\documentclass[conference]{IEEEtran}
\IEEEoverridecommandlockouts
\usepackage{amsmath,amssymb,amsfonts}
\usepackage{algorithmic}
\usepackage{graphicx}
\usepackage[inline]{enumitem}

\usepackage{textcomp}
\usepackage{xcolor}

\usepackage{lipsum}
\usepackage{multicol}
\usepackage{soul}
\usepackage{hyperref}
\usepackage[disable]{todonotes}
\usepackage[sort&compress,numbers]{natbib}

\usepackage{setspace}
\usepackage{comment}

\usepackage[ruled,vlined,linesnumbered]{algorithm2e}

\usepackage{subfig}
\usepackage{array}
\usepackage{longtable}

\usepackage{booktabs}

\def\BibTeX{{\rm B\kern-.05em{\sc i\kern-.025em b}\kern-.08em
    T\kern-.1667em\lower.7ex\hbox{E}\kern-.125emX}}

\usepackage{fancyhdr} % see https://tex.stackexchange.com/questions/64667/text-in-footer-on-first-page-but-no-page-numbers
\begin{document}

% See references.bib to control maximum number of authors: https://tex.stackexchange.com/questions/164017/limiting-the-number-of-authors-in-the-references-with-ieeetran?noredirect=1&lq=1
\bstctlcite{IEEEexample:BSTcontrol}

\title{Provenance Data in the Machine Learning Lifecycle in Computational Science and Engineering}

\author{\IEEEauthorblockN{
Renan Souza\textsuperscript{\S,o}, 
Leonardo Azevedo\textsuperscript{\S},
V\'{i}tor Louren\c{c}o\textsuperscript{\S},
Elton Soares\textsuperscript{\S},
\\
Raphael Thiago\textsuperscript{\S},
Rafael Brand\~{a}o\textsuperscript{\S},
Daniel Civitarese\textsuperscript{\S},
Emilio Vital~Brazil\textsuperscript{\S},
\\
Marcio Moreno\textsuperscript{\S},
Patrick Valduriez\textsuperscript{\#},
Marta Mattoso\textsuperscript{o},
Renato Cerqueira\textsuperscript{\S},
Marco A. S. Netto\textsuperscript{\S}
}
\IEEEauthorblockA
{
\textsuperscript{\S}IBM Research, \textsuperscript{o}Federal Univ. of Rio de Janeiro, \textsuperscript{\#}Inria \& LIRMM, U. Montpellier
}
}

\maketitle

%%% MY COMMANDS

\newcommand{\needref}[1][?]{\colorbox{lightgray}{[R #1]}}
% Usage: \needref{} or \needref[any hint on the refs]{}

\newcommand{\eg}{\emph{e.g.},}
\newcommand{\egUpper}{\emph{E.g.},}
\newcommand{\ie}{\emph{i.e.},}
\newcommand{\etal}{\textit{et al.}}

\newcommand{\codebackground}[1]{\colorbox{black!5}{\parbox{\dimexpr\linewidth-2\fboxsep}{\fontfamily{pcr}\scriptsize#1}}}

\newcommand{\codefont}[1]{{\fontfamily{pcr}{\small{#1}}}}

% Counter
\newcounter{qcounter} 
\newcommand{\createQ}[2]{
    \refstepcounter{qcounter} \label{#1} \textit{\textbf{Q\ref{#1}:} {#2}}
}
\newcommand{\refQ}[1]{Q#1}
% Usage: first create the reference using \createQ{label_name}, then you can make a reference to it using refQ{label_name}

\newcommand{\MLCycle}{ML lifecycle in CSE}
\newcommand{\queries}{\hyperref[tab:queries]{Q1--Q6}}
\newcommand{\textonto}[1]{\texttt{\small{#1}}}
\newcommand{\textontohead}[1]{\textbf{\emph{#1}}}
\newcommand{\textsoftware}[1]{\textsc{#1}}

\newcommand{\alert}[1]{\textcolor{red}{#1}}

\newcommand{\provcapturesystems}{\cite{komadu_suriarachchi_big_2018, lucas_carvalho2018provcompliant, joao_survey_2019, silva_raw_2017, silva_capturing_2018}}

\begin{abstract}
 Machine Learning (ML) has become essential in several industries. In Computational Science and Engineering (CSE), the complexity of the ML lifecycle comes from the large variety of data, scientists' expertise, tools, and workflows. If data are not tracked properly during the lifecycle, it becomes unfeasible to recreate a ML model from scratch or to explain to stakeholders how it was created. The main limitation of provenance tracking solutions is that they cannot cope with provenance capture and integration of domain and ML data processed in the multiple workflows in the lifecycle, while keeping the provenance capture overhead low. To handle this problem, in this paper we contribute with a detailed characterization of provenance data in the \MLCycle{}; a new provenance data representation, called PROV-ML, built on top of W3C PROV and ML Schema; and extensions to a system that tracks provenance from multiple workflows to address the characteristics of ML and CSE, and to allow for provenance queries with a standard vocabulary. We show a practical use in a real case in the O\&G industry, along with its evaluation using 48 GPUs in parallel.
\end{abstract}


\thispagestyle{fancy}%
\section{Introduction} \label{sec_intro}

Machine Learning (ML) has been fundamentally changing Computational Science and Engineering (CSE) in various ways~\cite{MLCSE}. 
Techniques, such as statistical relational learning and deep learning, have been used to extract knowledge from data, with application domains ranging from Computational Physics to Agriculture and Oil and Gas (O\&G)~\cite{gil_intelligent_2018, raissi_pinns_2019, rodrigues2018deepdownscale,salles19}. 
Obtaining reliable ML models involves a complex ML lifecycle~\cite{MLLifeCycle}, 
which is critical in large-scale CSE projects~\cite{gil_intelligent_2018, RoleMLinWorkflows}.
The \textit{\MLCycle{}} depends on transforming raw data into trained models, which requires multiple, distributed workflows that use a wide variety of algorithms, data, data processing tools and data stores; demands execution in machines ranging from standalone servers to cloud or HPC clusters; and is carried out by multidisciplinary teams, including domain scientists, computational scientists and engineers, and ML specialists. Given the heterogeneous nature of the lifecycle, it is difficult to track, in an integrated way, the data transformations that occur throughout the lifecycle while keeping the execution overhead low, which is a major concern among CSE users. In practice, tracking the data in the data transformations is often done manually, which is time consuming and error prone. This is problematic for several reasons, ranging from scientific (\eg{} jeopardizes reproducibility) to business (\eg{} users may be less likely to apply a trained model, even with best performance, if they do not understand the transformations in the lifecycle).

% As further detailed in this paper, the \textit{\MLCycle{}} can be divided into three major phases: 
% \textit{data curation}, 
% \textit{learning data preparation}, 
% and \textit{learning}. 

Data lineage (\ie{} data provenance) helps reproducing, tracing, assessing, and understanding data and their transformation processes \cite{herschel_survey_2017}. 
Solutions for provenance data tracking for ML have been proposed~\cite{miao_towards_2017, zhang_diagnosing_2017, zaharia_accelerating_2018}, but with focus on learning phases only, thus limiting an integrated view of domain-specific data, processed in the early phases of the lifecycle, with ML data.
Besides, users need to migrate their workflows to a different software ecosystem or change the way they develop, which may compromise adoption in CSE.
Another approach is to add provenance tracking to workflows, reducing the need to change the development practice \provcapturesystems{}. Nonetheless, solutions following this approach do not support the lifecycle, which requires three main capabilities: provenance tracking in multiple workflows that use heterogeneous data and stores; a provenance data representation with ML-specific vocabulary; and providing for integrated data analysis through provenance while keeping the capture overhead low.
Another solution following this approach is ProvLake \cite{provlake_escience_2019}, which has low capture overhead in multiple workflows that use heterogeneous data stores. However, similarly to the other solutions, its provenance data representation is based on W3C PROV only, thus lacking ML-specific vocabulary, limiting its support for the \MLCycle{}. Allowing for such provenance analysis that integrates both ML data and domain-specific data while keeping capture overhead low in HPC workflows is hard.

In this paper, we propose an end-to-end solution for tracking data transformations that occur in the \MLCycle{}, 
from the curation of raw data until the generation of trained models, by providing efficient provenance capture and data analysis through provenance queries.
Our approach is to model the lifecycle as multiple workflows interconnected with data and to track provenance as the workflows execute. By adding provenance capture calls in the workflows, users can perform ML monitoring (\eg{} the evolution of model performance as the training iterates) and more comprehensive provenance analyses that join domain-specific data with ML data generated in the lifecycle. The main contributions of the paper are:
\begin{enumerate}[label=(\roman*)]
    \item a characterization of provenance data in the \MLCycle{} (Sec.~\ref{sec_ml_lifecycle});
    \smallskip
    \item PROV-ML: a new data representation, which combines W3C PROV~\cite{W3CPROV} with W3C ML Schema~\cite{publio2018}, for provenance of the \MLCycle{} (Sec.~\ref{sec_datarepresentation});
    \smallskip
    \item  extensions of the ProvLake~\cite{provlake_escience_2019} to support provenance tracking and analysis following the PROV-ML (Sec.~\ref{sec_provlakeml}). 
    Experiments show its practical use in a real O\&G case in a testbed of 48 GPUs (Sec.~\ref{sec_exps}).

\end{enumerate}

%ection~\ref{sec_related_work} has related work and Section \ref{sec_conclusion} concludes.

\section{Workflow Provenance for ML in CSE} \label{sec_ml_lifecycle}
%There are several open challenges to be addressed for a more efficient and effective data pipeline management in the ML lifecycle \cite{MLLifeCycle}. In particular, in CSE, there are even further problems, mostly because of the complex nature of the scientific data and the exploratory nature of science. 
This work focuses on provenance in workflows formed by chained data transformations composing the \MLCycle{}, aiming at supporting the data analysis in a large-scale CSE project.
Before we characterize the data analysis through provenance (Sec. \ref{subsec_provanalysis}), we first characterize the lifecycle's personas (Sec. \ref{subsec_users}) to provide for analysis addressing the users' needs, then we describe the lifecycle (Sec. \ref{subsec_mllifecycle}).

\subsection{Personas in the ML Lifecycle in CSE} \label{subsec_users}

\begin{figure}[!h]
  \includegraphics[width=\linewidth]{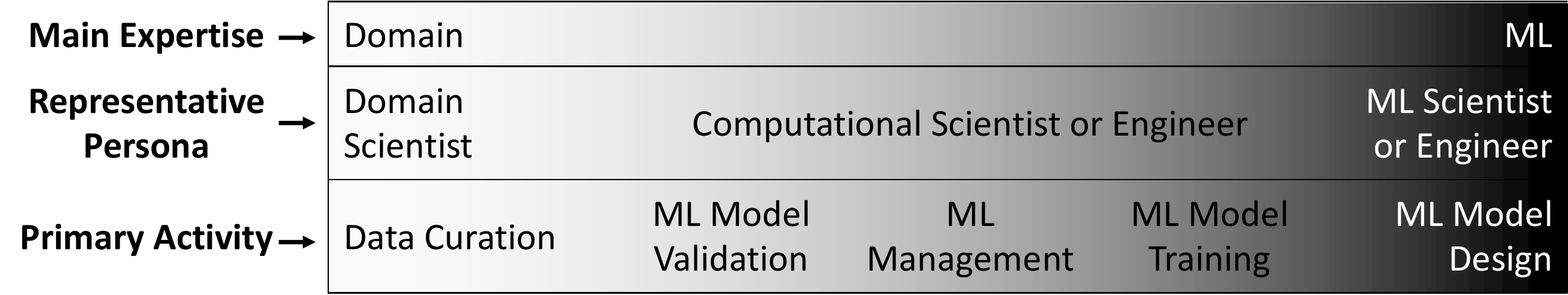}
  \caption{Spectrum of expertise and personas in the lifecycle.}
   \label{fig:personas}
\end{figure}

Large-scale CSE projects are often multidisciplinary, with collaborating users with different skills on the domain data, \eg{} mathematics, physics, statistics, computational methods, and ML. 
These users perform distinct types of analysis and have different provenance requirements. In order to position the personas and their primary activities in the~\MLCycle{}, we adapt background work on traditional scientific workflows ~\cite{souza_datareduction} and ML \cite{MLLifeCycle}.
Fig. \ref{fig:personas} illustrates how expertise, representative personas, and primary activities fall under an expertise spectrum ranging from scientific-domain only (fully white on the left) to ML only (fully black on the right).

\smallskip
\noindent \textbf{Domain scientists.} They have in-depth knowledge of the domain data and use specialized tools to interpret, visualize, and clean the scientific data~\cite{silva_capturing_2018}, thus 
playing an important role to curate the raw scientific data, specify domain matters, and validate results. Examples are geoscientists, agronomists, and experimental physicists. 
They can validate ML models qualitatively, \ie{} they can check if results are reasonable given the characteristics of the domain.
Oftentimes, such validations contradict the numeric results obtained by ML algorithms.
They are also paramount in hypothesis definition and in verifying simplifications made about domain's problems.
They contribute by collecting domain-specific annotations from technical reports and articles, and link such annotations to the raw scientific data, augmenting the possibilities for enhanced analyses, contributing to the training. They are critical to providing labeled scientific data to supervised learning algorithms. 

\smallskip
\noindent
\textbf{Computational scientists and engineers.} They have high computational skills, often with abilities to develop parallel scripts and execute them in HPC clusters.
Examples are computational physicists, engineers. 
They are highly knowledgeable in the domain, although not as in-depth as the domain scientists. 
They are familiar with traditional numerical simulations that require HPC, which need complex scientific data analyses to guide the fine-tuning of parameters \cite{souza_keeping_2019, silva_capturing_2018, MLCSE}. 
In the lifecycle, they are often the ML model trainers, who tune parameters, a very usual task when training ML models. 
They use their knowledge on the domain to make decisions, \eg{} to filter relevant parts of the training datasets that are guaranteed to respect the physical constraints of the problem. Some users with more in-depth knowledge of ML techniques, 
\ie{} those who are more towards the black portion of the spectrum in Fig.~\ref{fig:personas}, can design new ML models. 
They can, for instance, design Physically-informed Neural Networks (PINNs)~\cite{raissi_pinns_2019}, which embed domain-specific physical constraints in the ML models.
These users can be responsible for validating the ML model 
and, more experienced users with considerable ML and domain knowledge, help in the overall analyses of the produced models, their quality, how they were used, etc.

\smallskip
\noindent
\textbf{ML scientists and engineers.} They have in-depth knowledge of statistics, ML algorithms, and software engineering. 
They design new ML models and develop scripts typically using ML libraries like TensorFlow, PyTorch, and Scikit Learn.
They are familiar with software engineering techniques (\eg{} continuous integration, test-driven development, cloud deployment) 
and can use different kinds of DBMSs to store data. 
They often train the ML models they design, often in HPC clusters. 
Moreover, to be able to develop effective models, they also have some domain knowledge.

\smallskip
\noindent
\textbf{Provenance specialists.} In addition to those three main personas, Provenance specialists play an essential role in a CSE project by managing data provenance in the lifecycle. 
They design the provenance schema for applications and guide other users to add provenance capture calls to the workflows. 
Thus, they need knowledge in the scientific domain and ML.  
They also support other personas to analyze provenance, domain-specific, execution, and ML data.

\subsection{The \MLCycle{}} \label{subsec_mllifecycle}

\begin{figure}[!h]
  \centering
  \includegraphics[width=\columnwidth]{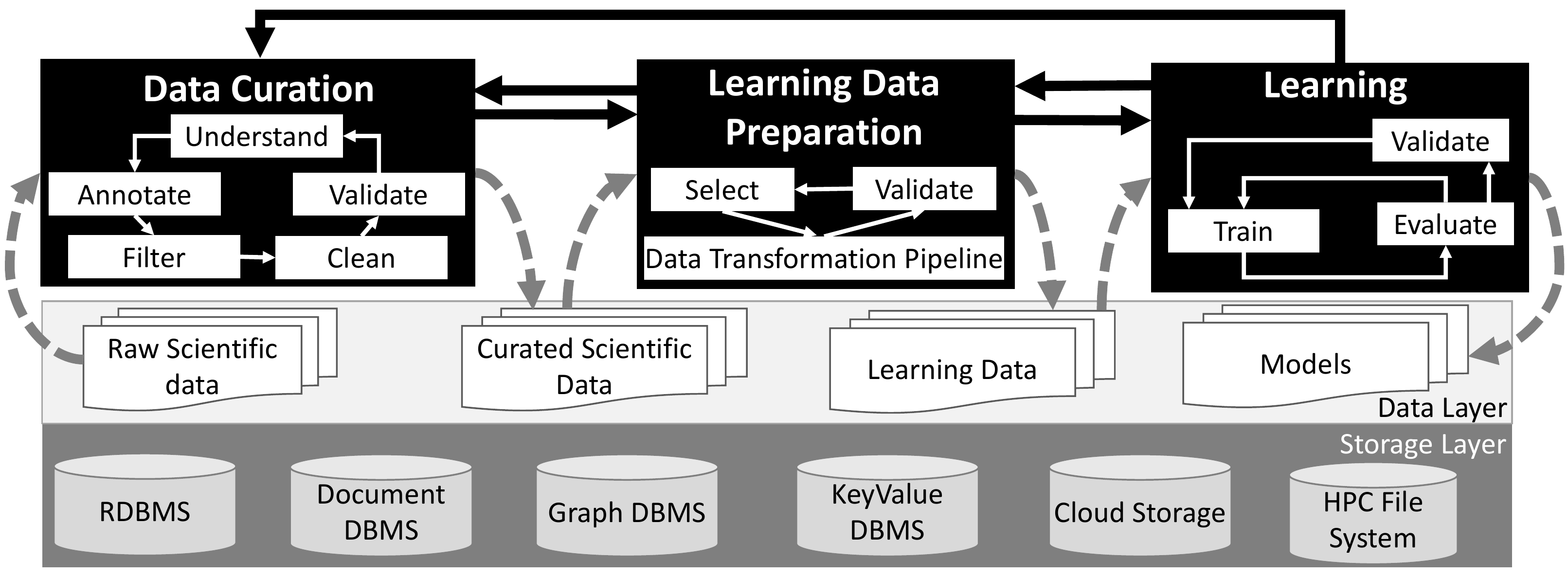}
  \caption{The ML lifecycle in CSE.}
   \label{fig:lifecycle}
\end{figure}

We can divide the \MLCycle{} into three major phases: \textit{data curation}, \textit{learning data preparation}, and \textit{learning} (Fig.~\ref{fig:lifecycle} --- dashed arrows are data flows and solid arrows are interactions between phases). Our view of the lifecycle is inspired by Polyzotis \etal{}' survey \cite{MLLifeCycle}. 
Although they proposed a generic view, which can be applied to CSE, there remains the need for a focused view on the problems inherent to CSE. 
We grouped the inner phases into major phases, organizing the activities according to the scientific data manipulated and the personas involved in the major phases. 

% \todo[inline]{LGA: sugiro alterar para "So, we grouped phases into macro phases that reflect ML lifecycle, \ie{} we organized the activities according to the scientific data manipulated and the CSE personas that manipulate them in each major phase."}

%%%%%%------------------------------------------
%%%%%%------------------------------------------
\smallskip
\noindent \textbf{Data curation.} 
It is the most complex phase of the lifecycle, especially because of the nature of the scientific data. 
To achieve automated knowledge extraction from scientific data promoted by ML, much manual and highly specialized work is performed by the users (mainly domain scientists). 
There is a huge gap between raw scientific data and useful data for consumption (\eg{} data to serve as input to train ML models). 
Datasets are typically large, up to terabytes in a single file. They may contain geospatial-temporal data, stored as huge matrices in well-known scientific formats, like HDF5. 
Also, some files are stored using domain-specific formats, \eg{} SEG-Y format for seismic data, widely used in the O\&G industry. 
Specialized formats in CSE domains may require industry-specific software and domain-specific knowledge to inspect, visualize, and understand the data. 
%Frase removida para ganhar espaço: "Moreover, raw data are frequently noisy and problematic."
%Frase removida para ganhar espaço:Using the seismic data example, SEG-Y files are known for being a ``standard without standards'' in the O\&G industry, as many geoscientists do not follow the recommendations to generate a SEG-Y file properly, thus missing data is very common. %
In addition, users can use metadata and textual reports to annotate the data with extra domain-specific knowledge, without which would be nearly impossible to make the data useful for ML algorithms. 
%Frase removida para ganhar espaço: "In this phase, domain scientists can also label the data for supervised algorithms and extend knowledge about the raw data by associating, in a knowledge database, pieces of texts, for example, extracted from documents like academic papers, technical reports, or surveys."}

Considering the heterogeneous nature of the data, ``it is unreasonable to assume that data lives in a single source'' (\eg{} a single file system or DBMS) \cite{MLLifeCycle}. 
%Frase removida para ganhar espaço: "In fact, data are often spread across multiple data stores."
For instance, raw files can be stored in file systems or cloud stores, domain-specific annotations can be stored in a Semantic Graph DBMS (\eg{} Triple Store) with domain ontologies, and curated data can be stored in a NoSQL DBMS, such as Document DBMSs. 
Then, computational scientists and engineers write data-intensive workflow scripts to clean, filter, and validate the data. For instance, they check if the geolocalization of the data files is consistent.
These scripts transform the raw data into curated data by consuming and generating data from those data stores. 
Each of these inner phases inside the data curation phase is highly interactive, manual, and may execute independently.
In other words, users may run different scripts to execute these phases, several times, in an \textit{ad-hoc} way and any order. Also, they run these scripts in different machines, such as in an HPC cluster or in the cloud, or even on the users' desktop. These phases occur in a cycle, which stops when the users consider the data ``curated''. 
These curated data are significantly more organized and easier to analyze and understand. 
%Removido para ganhar espaço: by the users in the CSE project. 
In the context of ML, it is ready to be transformed into training data.

%%%%%%------------------------------------------
%%%%%%------------------------------------------

\todo[inline,color=orange]{Daniel+Emilio: completar este paragrafo abaixo}
\smallskip
\noindent \textbf{Learning data preparation.} Model trainers select relevant parts of the curated data to be used for learning. 
For instance, if the ML task is to classify geological structures~\cite{salles19}, seismic images will need to be correlated with annotations -seismic interpretation-, creating annotated samples. 
After selecting the data, model designers develop scripts that transform the data into training datasets. Typical transformations include image crop, quantization, scale, among others. 
In this phase, users frequently use domain-specific libraries to manipulate raw scientific data. 
Due to data complexity, oftentimes data need to be manually inspected before it can be used as input for the learning phase.
%%%%%%------------------------------------------
%%%%%%------------------------------------------

% Daniel
\todo[inline,color=orange]{Daniel: falar o que significa Validate no paragrafo abaixo e checar a parte em amarelo (that will be optmized...)}
\smallskip
\noindent \textbf{Learning.} In this phase, model trainers select the input training datasets, optionally they choose validation datasets, and choose training parameters (\eg{} in deep learning they can choose ranges of epochs and learning rates) that will be optimized in the training process. Trainers can use their domain knowledge to discard input training datasets that will unlikely provide good results. The training process is compute-intensive, typically executed as a job submitted in an HPC cluster. 
One single training process often generates multiple trained models, among which one is chosen as the ``best'' depending on evaluation metrics (\eg{} MSE, accuracy, or any other user-defined metric). 
Moreover, as the training process takes a long time, trainers need to monitor it by, \eg{} inspecting how the evaluation metrics are evolving while the training process iterates. 
They can wait until completion or interrupt the training process, change parameters, re-submit the training in an iterative way until satisfied with results. 
%Elaborei a frase anterir para tentar reduzir espaço: If they do not like the results, they can either specify different input training datasets, parameter values, or evaluation metrics and re-submit a new training process. This happens in a cycle until the model trainer is satisfied with the results. 

\subsection{Characterizing Provenance Analysis in ML for CSE} 
\label{subsec_provanalysis}

Provenance data in workflows contain a structured record of the data derivation paths within chained data transformations, along with the parameterization of each transformation \cite{silva_raw_2017, souza_keeping_2019}. 
Provenance data are usually represented as a directed graph where: vertices are instances of entities (typically data) or activities (typically the data transformations) or agents (typically the users); and, edges are instances of relationships between vertices \cite{W3CPROV}. 
Scientists use provenance data for reproducibility and result understanding \cite{herschel_survey_2017}. This kind of provenance consumption, which often occurs \textit{post mortem}, \ie{} after workflow execution, is characterized as offline provenance analysis. 
%In addition to this characteristic of analysis, there are many other ways to leverage provenance analysis for the workflows of the \MLCycle{}, and characterizing them is our goal in this section. 
A characterization of provenance analysis to leverage ML in support of workflows is surveyed by Deelman \etal{} \cite{RoleMLinWorkflows}. 
We propose here a taxonomy to classify provenance analysis in support of ML, by considering three classes: \textit{data}, \textit{execution timing}, and \textit{training timing}. 
We provide the characterization based on the data being analyzed, using query examples (listed in Table \ref{tab:queries}) in our use case in O\&G.

\begin{table}[!t] 
\caption{Examples of provenance queries in ML for CSE.}
\footnotesize
\begin{tabular}{>{\centering\arraybackslash}m{0.005cm}m{7.7cm}}
% \hline
% \hline
\toprule
\multicolumn{1}{c}
{Q1} & Given a trained model, what are the geographic coordinates, oil basin and field, and the number of seismic slices of the seismic in the training dataset?
\\ \hline
Q2 & Given a trained model, what is the tile size, the noise filter threshold, and the ranges of seismic slices that were selected to generate the training set used to adjust this model?
\\ \hline
Q3 & Given a training set, what are the values for all hyperparameters and the evaluation measure values associated with the trained model with least loss?
\\ \hline
Q4 & What are the average, min, and max execution times of each batch iteration inside each epoch of the deep neural network training, given a training dataset? 
\\ \hline
Q5 & What is the execution time on average per batch iteration, per epoch, and what are the evaluation metrics of the trained models that used the training dataset generated for a given range of seismic slices? 
\\ \hline
Q6 & Given the training dataset used in Q5, what was the seismic data file used, along with its number of slices, related oil basin, and field?                           \\
% \hline
% \hline
\toprule
\end{tabular}
\label{tab:queries}
\end{table}

\smallskip
\noindent
\textbf{Use case.} The use case addresses seismic surveys, which are indirect measures of the earth subsurface that can be organized into slices (images).
These surveys cover hundreds of square kilometers and help to interpret the geology and find possible hydrocarbons accumulations. 
The seismic data have a very complex workflow and can suffer from many problems, like noise and shadows (regions with low signal). 
Also, the geological structures vary from point to point in the earth, imposing significant challenges to the ML algorithms. Next, we characterize the data involved in the lifecycle.

% \begin{figure}
%   \includegraphics[width=\linewidth]{img/taxonomy.pdf}
%   \caption{A taxonomy for ML provenance analysis in CSE.}
%   \label{fig:provml_taxonomy}
% \end{figure}

\smallskip
\noindent \textbf{Data} class includes \textit{domain-specific}, \textit{machine learning}, and \textit{execution}.
Provenance data may be augmented with these data, increasing the scope of provenance analysis.
%The provenance analysis can be hybrid, \ie{}, include more than a data class in the same analysis. 

\textit{Domain-specific data} are the main data processed in the data curation phase (Sec. \ref{subsec_mllifecycle}).
Approaches to add domain data into provenance analysis include, \eg{} raw data extraction \cite{silva_raw_2017} and utilization of  semantic domain databases associated to provenance databases \cite{provlake_escience_2019}. 
For raw data extraction, quantities of interest are extracted from large raw data files, and for domain databases, domain scientists may provide relevant information and metadata about the raw data and store them in knowledge data graphs (\eg{} in Triple Stores).
%Such domain-specific knowledge can be stored as knowledge data graphs in Triple Stores with domain-specific ontologies, and also be related to provenance data while the workflows run, augmenting the provenance data analysis with domain-specific data. 

% \createQ{q:q1}{Given a trained model, what is the geographic coordinates, the oil basin, oil field, and number of seismic slices of the seismic data used as training dataset?}

\textit{Machine learning data} include training data and generated trained models, which are more related to the learning data preparation (\eg{} Q1) and learning (\eg{} Q2 and Q3) phases (Fig.~\ref{fig:lifecycle}). These queries exemplify that the parametrization within the data transformations and relevant metadata of the generated data (both training data and trained model) are important for provenance analysis. 
%For example, for the Training Data Generation phase:

% \createQ{q:q2}{Given a trained model, what is the tile size, the specified noise threshold to reject too noisy tiles, and ranges of seismic slices that were selected to generate the training dataset used to train this model?}

% %For the Training phase:

% \createQ{q:q3}{Given a training dataset, what are the values for all hyperparameters and the evaluation measure values associated to the trained model with least loss?} 

\textit{Execution data.} Besides model performance metrics (\eg{} accuracy), users need to assess execution time and resources consumption of their workflows. They need to inspect if a critical block in their workflow (\eg{} the one that demands high parallelism) is taking longer than usual or if other parts are consuming more memory than expected. For this, provenance systems can capture system performance metrics and timestamps (\eg{} Q4). Metadata, such as data store metadata (\eg{} host address), HPC cluster name and nodes in use, can be captured and associated with the provenance of the data transformations for extended analysis.

% \createQ{q:q4}{What are the average, min, and max execution times of each iteration inside each epoch of the deep neural network training, given a training dataset?}

\textit{Hybrid.} Users can combine these data. For instance, in Q5, the analysis queries data processed in workflows in the learning data preparation and learning phases, whereas Q6 uses the same data generated in the learning data preparation to analyze the raw files curated in the data curation phase.

\smallskip
\noindent \textbf{Execution timing} refers to if the analysis is done \textit{online}, \ie{} while at least a workflow is running, or \textit{offline}. 

\textit{Offline analysis.} The typical use of offline provenance analysis is to support reproducibility and historical results understanding, \eg{} understand the data curation phase of raw scientific files and relate with the generated trained ML models. 
The queries \queries{} can be executed offline.

\textit{Online analysis.} Users can use online provenance analysis to monitor, debug or inspect the data transformations while they are still running (\eg{} see the status, see how the intermediate results are evolving as the input parameters vary). 
%It is very useful, for instance, to monitor the model performance metrics while the training iterates. 
The problem of adding low provenance data capture overhead is more challenging for provenance systems that allow for online analysis \cite{provlake_escience_2019}. 
Queries Q3--Q5 exemplify queries that can be executed online, \eg{} while a training process is running.

\smallskip
\noindent \textbf{Training timing} refers to whether the analysis performs
\textit{intra-training}---\ie{} to inspect one training process, \eg{} a training job running on an HPC cluster, or 
\textit{inter-training}---\ie{} analyses comprehending results of several training processes. 

\textit{Intra-training}. 
In an offline intra-training analysis, users are interested in understanding how well trained models generated in a given training process perform. All queries, \queries{}, could be executed either online or offline, but Q3 and Q4 are more likely to be performed as online intra-training analysis.
%For such, all queries \queries{} can be used for a deep analysis of a training process executed in the Training phase, along with its inputs and outputs. 
%Online intra-training analyses are especially important for the Model Training activity because trainers need to monitor and debug how a specific training process is evolving. 
% Although \queries{} could also be executed online for deeper analysis, Q3 and Q4 are more likely to be performed as online intra-training analysis.

\textit{Inter-training}. This analysis refers to comprehensive queries to understand multiple training processes, \eg{} 
how each of them performed, 
which training datasets were used, 
how the training processes were parameterized. 
This is very important in the lifecycle, as it supports activities like Model Validation, Management, Training, and Design. 
Usually, they are used offline, but may also be performed online.
%With respect to execution timing, this is mostly used as offline analysis, though it could be used online to provide further information to model trainers who need to compare a running training process with other processes already finished. 
Queries \queries{} fit this class when analyzing multiple trained models generated in different training processes.

\smallskip
\noindent \textbf{Further characterization.} Other classes worth mentioning for provenance analysis for ML in CSE are: \textit{data store}---data are distributed onto multiple stores, like file systems, cloud stores (\eg{} IBM Cloud Object Storage), Relational or NoSQL DBMSs \cite{provlake_escience_2019}; \textit{provenance data granularity}---provenance of files (\ie{} references to files consumed and generated in a script), functions calls (arguments and outputs), blocks of code, and stack traces \cite{joao_survey_2019}; and \textit{provenance analysis direction: \textit{forward} or \textit{backward}}---generally, forward queries analyze from raw scientific files or training datasets to trained models (\eg{} Q3--Q5), whereas backward queries analyze from trained models to training datasets or raw files (\eg{} Q1, Q2, Q6).

\section{ML Provenance Data Representation} \label{sec_datarepresentation}

\begin{figure*}[t]
   \centering
     %trim={<left> <lower> <right> <upper>}
   %\includegraphics[clip,trim={1.0cm 21cm 1cm 1cm},width=\linewidth]{img/PROV-Lake-ML-Model.pdf}
\includegraphics[clip,trim={0.0cm 10.6cm 0.0cm 10.0cm},width=\linewidth,keepaspectratio]{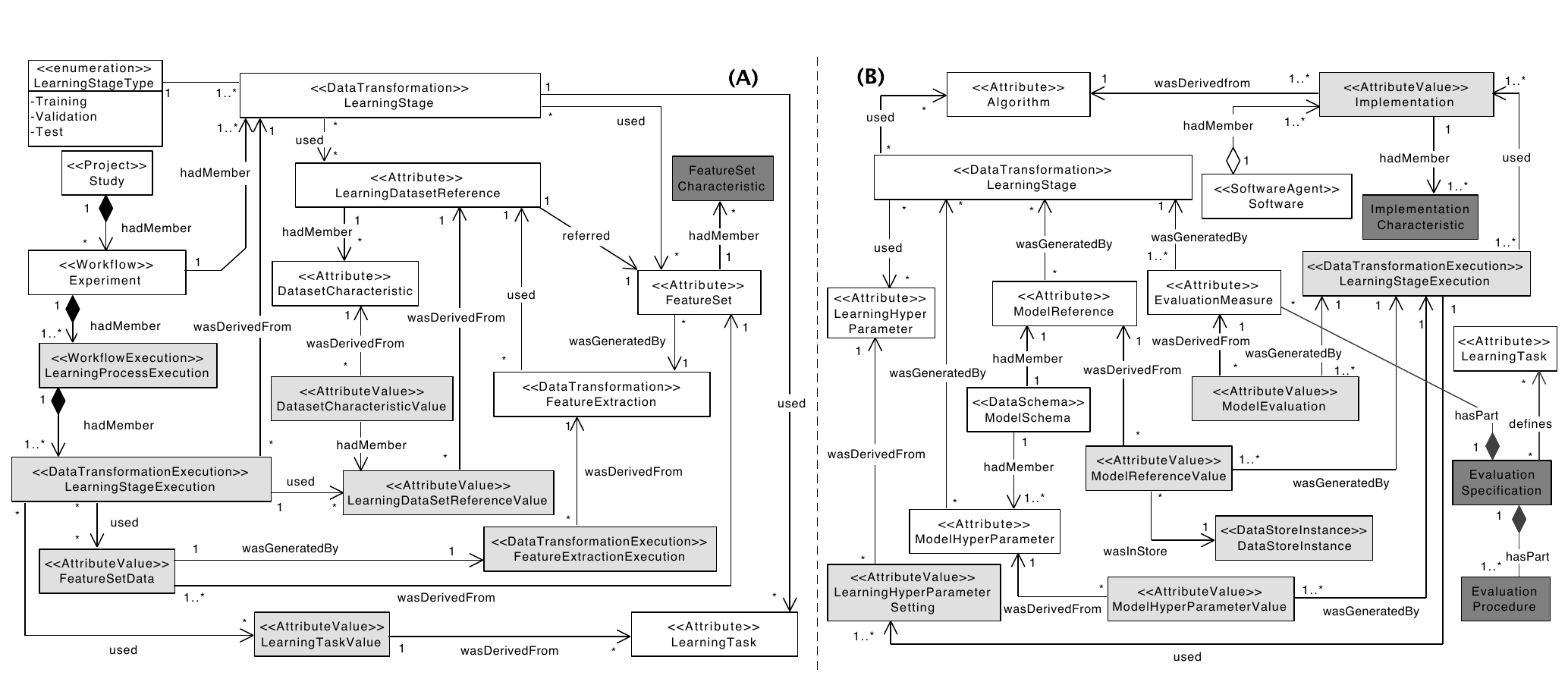}
   \caption{PROV-ML: a W3C PROV- and W3C ML Schema-compliant workflow provenance data representation.}
   \label{fig:prov-lake-ml-model-data}
   \vspace{-3mm}
\end{figure*}

There are many workflow provenance tracking solutions ~\cite{komadu_suriarachchi_big_2018, lucas_carvalho2018provcompliant, silva_raw_2017, provlake_escience_2019}, but they are often based on W3C PROV~\cite{W3CPROV} (and extensions) only. 
Thus, they are too generic in terms of provenance data representation and analysis, which makes the adoption for ML more difficult. 
An existing work on ML data representation is the W3C ML~Schema~(MLS)~\cite{W3CML}. Although the MLS has some provenance representation, a MLS-based only representation does not meet the needs either. It does not have a clear distinction between prospective and retrospective provenance data, which compromises query capabilities because prospective provenance provides the abstraction layer to specify provenance analysis over data generated in workflows' execution (\ie{} retrospective provenance). Also, MLS does not separate the learning stages (training, validation and test), which would enable finer analysis based on specific stages. Finally, it is not designed to allow for representation of domain-specific data generated at early phases of the lifecycle (\eg{} data curation).

To address these problems, this section introduces PROV-ML. To the best of our knowledge, it is the first provenance data representation for workflows in the \MLCycle{}. 
It is compliant with both W3C PROV and MLS, and extends ProvLake's workflow provenance data representation~\cite{provlake_web}, which is an extension of PROV.

PROV-ML provides detailed support for the learning and learning data preparation phases of lifecycle. 
It inherits the benefits of ProvLake, enabling the integration of provenance of domain-specific data processed by workflows in the curation phase.
PROV-ML is depicted in Fig.~\ref{fig:prov-lake-ml-model-data}, where Fig.~\ref{fig:prov-lake-ml-model-data}(A) shows the relation of the learning phase with the input data and the goal of a ML workflow (\ie{} a workflow in the learning phase); and Fig.~\ref{fig:prov-lake-ml-model-data}(B) represents the relation of the learning phase with its technique and parameters. Classes in white background represent prospective provenance; light gray, retrospective; and dark gray represents specific concepts inherited, as is, from MLS. PROV-ML classes are described on Table~\ref{tab:provlakeml-classes}. Further details on PROV-ML are online \cite{provlake_web}.
% V4
\begin{table}[!h]
\footnotesize
\caption{PROV-ML data representation classes.}
\begin{tabular}{>{\centering\arraybackslash}m{3.25cm}m{4.5cm}}
% \hline
% \hline
\toprule
\textbf{Class} & \textbf{Description} \\
\hline
Study & Investigation (\eg{} research hypothesis) leading ML workflow definitions.
\\ \hline
Experiment & The set of analyses (\eg{} research questions), that drives the ML workflow. 
\\ \hline
LearningProcessExecution & An ML workflow execution. This is equivalent to \textit{mls:Run} and was renamed to explicitly preserve the aspects of retrospective provenance, which are not explicitly handled in MLS. 
\\ \hline
LearningTask and LearningTaskValue & Defines the goal of a learning process, \ie{} the ML task  (\eg{} \textit{LearningTask}: \textit{Classification}; \textit{LearningTaskValue}: \textit{Seismic Stratigraphic Classification}). 
\\ \hline
LearningStageType & A stage in the learning process. It is one of: training, testing or validation.
\\ \hline
LearningDatasetReference & Defines the dataset to be used by a \textit{LearningStage} and \textit{LearningDatasetReferenceValue} is the dataset reference used in a \textit{LearningStageExectution}.                                                                                                               \\
\hline
DatasetCharacteristic and DatasetCharacteristcValue & Defines metadata about the \textit{LearningDatasetReference} (\eg{} \#instances),  and \textit{DatasetCharacteristcValue} relates with a \textit{LearningDatasetReferenceValue} (\eg{} \#instances =8).                      \\ \hline
FeatureSet and FeatureSetData & Defines the features \textit{FeatureExtraction} to generate over  \textit{LearningDatasetReference} and, \textit{FeatureSetData} is the generated values in the execution.                                                                           \\ \hline
FeatureSetCharacteristic  & Defines the set of metadata that describes the \textit{FeatureSet} (\eg{} number of features, features' type).                                  \\ \hline
FeatureExtraction and FeatureExtractionExecution & Defines the features retrieval process.                                           \\ \hline
Software & Defines a collection of ML techniques' implementations (\eg{} Scikit-Learn).                                                          \\ \hline
Algorithm & ML technique with no associated technology, software or implementation (\eg{} k-means clustering technique).        \\ \hline
Implementation & Defines the retrospective aspect of an \textit{Algorithm}, \ie{} an ML technique's implementation in a software (\eg{} Scikit-Learn's k-means implementation).                                                                                                                   \\ \hline
ImplementationCharacteristic & Defines the implementation's set of metadata, (\eg{} version, git hash).                                       \\ \hline
LearningHyperParameter & Defines the prior parameter of an \textit{Algorithm} used by a \textit{LearningStage}.                                  \\ \hline
LearningHyperParameter Setting & Defines the parameter values of an execution (\eg{} the $k$ value in a k-means clustering technique, range of epochs in a neural network training).                                                                                                                                    \\ \hline

ModelSchema & The scope of the resulting model.                                                                                                  \\ \hline
ModelReference and ModelReferenceValue & The resulting model of a \textit{LearningStage} should generate and the generated value (\eg{} the trained model after the training stage).                                                                                                                  \\ \hline
ModelHyperParameter and ModelHyperParameterValue    &  Hyperparameters a \textit{LearningStage} generates and the resulting model with their values (\eg{} the epoch which the resulting model was generated), respectively.                                                \\ \hline
DataStoreInstance & Resulting model (\ie{} \textit{ModelReferencevalue}) storage.                                    \\ \hline
EvaluationMeasure and ModelEvaluation & A measure a \textit{LearningStage} should evaluate and its associated value generated in execution (\eg{} the precision of classifier model).                                                                                                      \\ \hline
EvaluationSpecification and EvaluationProcedure & Classes directly inherited from MLS, with their semantics preserved. \\
% \hline      
% \hline
\toprule
\end{tabular}
\label{tab:provlakeml-classes}
\end{table}

\section{ProvLake in the \MLCycle{}} \label{sec_provlakeml}

% \begin{figure*}
%   \centering
%   %\includegraphics[width=13cm,height=10cm,keepaspectratio]{img/architecture.pdf}
%   \includegraphics[width=0.50\linewidth]{img/architecture.pdf}
%   \caption{Architecture of ProvLake deployed onto two clusters: one for data curation and preparation and other for training.}
%   \label{fig:architecture}
%   \vspace{-3mm}
% \end{figure*}

To address provenance tracking and analysis throughout the \MLCycle{}, our approach is to model it as multiple workflows with chained data transformations, where the workflows are interconnected through data. Provenance tracking comprises provenance capture, the creation of the provenance relationships (\eg{} associations between the processes and the consumed and generated data), and storage of the provenance data. In our view, provenance tracking systems that can be coupled to workflows ~\cite{komadu_suriarachchi_big_2018, lucas_carvalho2018provcompliant, silva_raw_2017, provlake_escience_2019} provide the flexibility needed in large-scale CSE projects, as opposed to moving workflows' executions and data to be managed by a single orchestration system, like a Workflow Management System. Workflow provenance capture systems usually address scripts as workflows with chained functions, method, or library calls that execute data transformations, while capturing input arguments and output values from these calls. Among these solutions, ProvLake \cite{provlake_escience_2019} has been applied to capture provenance from multiple distributed workflows that consume and generate data from and to heterogeneous data stores, while keeping provenance capture overhead low. While these workflows execute, provenance data are captured and stored in a single provenance database, available for integrated analysis of the data generated throughout the lifecycle. This section describes ProvLake architecture and deployment in support of the lifecycle.
 
\begin{figure}[!h]
  \centering
  \includegraphics[width=\linewidth]{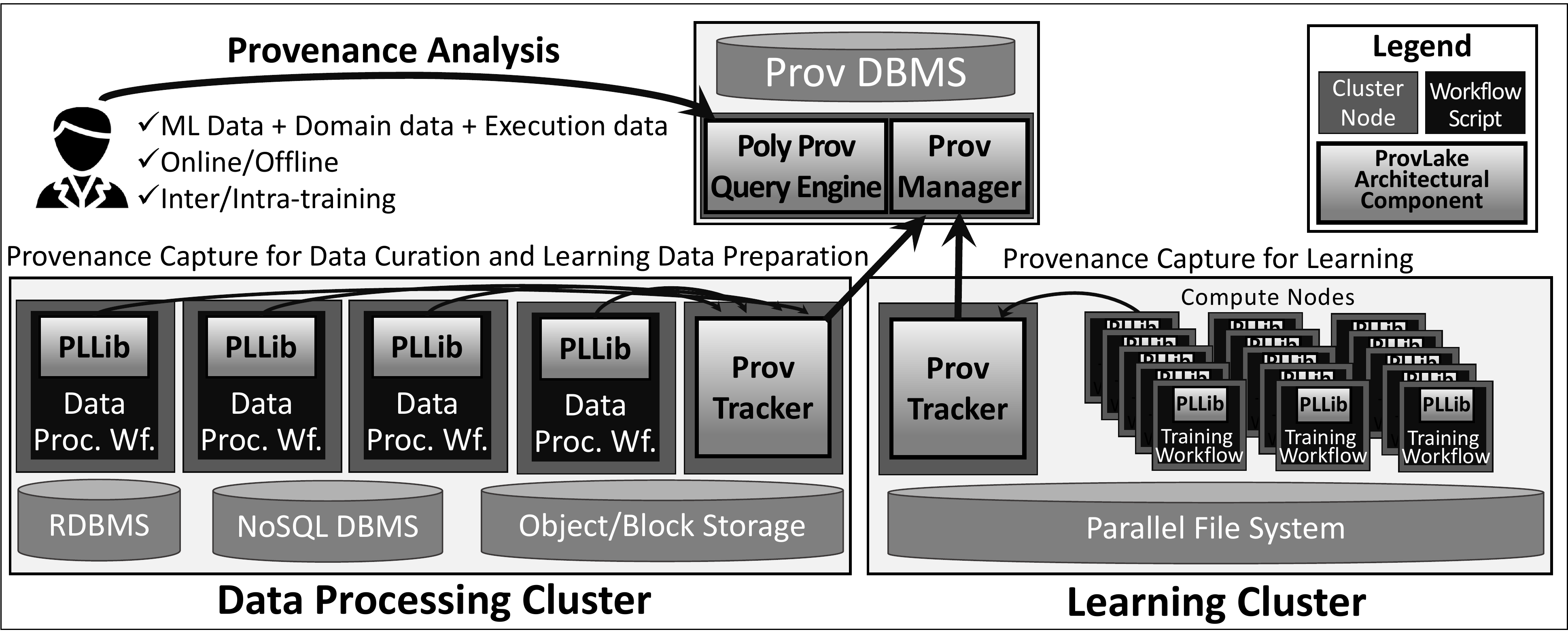}
  \caption{ProvLake architecture on an exemplary deployment on two clusters: for data preprocessing and for learning.}
  \label{fig:architecture}
  %\vspace{-3mm}
\end{figure}

\smallskip
\noindent \textbf{Architecture.} It has five main components (Fig. \ref{fig:architecture}):  
ProvLake Library (PLLib); 
ProvTracker; 
ProvManager; 
PolyProvQueryEngine;
and Prov DBMS (the DBMS that manages the provenance database). 

The workflows are instrumented with PLLib, imported as a library in the scripts, which is responsible for the provenance data capture.
In an offline manner and following the methodology described in a previous work \cite{provlake_escience_2019} to specify the workflows using prospective provenance data standards \cite{herschel_survey_2017}, users add provenance data capture calls using the PLLib. A provenance capture task happens when a data transformation executes, which typically occurs in a function call, in a  program execution, or in an iteration in an iterative workflow. 
%Scripts' developers define important regions in their code specifying the beginning and end of data transformation executions. 
As shown in a simplified pseudocode of a deep learning training (Algorithm~\ref{alg:alg1}), a provenance task is delimited within blocks of code in the scripts, illustrated with \codefont{prov.in()} and \codefont{prov.out()}, each generating a provenance capture event.

PLLib design has two goals: (i) to keep execution overhead low and (ii) to avoid major modifications in the user code while preserving the provenance data analytical capabilities. Because of the workflow specification using prospective provenance data, kept external to the workflow scripts and loaded only once at the beginning (Line 2 in Algorithm~\ref{alg:alg1}), the code modification and data to be sent to ProvTracker are reduced. 
Design principles such as queuing provenance requests, 
asynchronicity (\ie{} the workflow scripts do not wait for the provenance requests to be fully processed---the pipeline from the PLLib to ProvTracker, ProvManager, and Prov DBMS),
and reduction of system calls help reducing capture overhead \cite{provlake_escience_2019}. 
Provenance capture requests are queued and the maximum queue size is a configurable parameter. 
Moreover, users choose to store provenance data on disk only, rather than sending to ProvTracker, but in this case, online provenance analysis is not supported. 
Then, if disk only is not specified, when the scripts execute, provenance data are captured and sent to ProvTracker.

\begin{algorithm}[!b]
  \DontPrintSemicolon
  \begin{small}
  \KwIn{training hyperparameters, input data sets}
  \textit{import PLLib as prov}\;

  prov.init($prospective\_provenance$)\;
  \textit{...}\;
   prov.in($'training'$, $training\_hyperprms$, $input\_data\_references$)\;
   \For{$e = 1$ .. $max\_epochs$}{
      prov.in($'epoch'$, $e$)\;
      \textit{...}\;
      \For{$batch\_id$ in $data\_batches$ }{
          prov.in($'batch'$, $batch\_id$)\;
          \textit{...}\;
          prov.out($'batch'$, $loss\_value$)\;
       }
       prov.out($'epoch'$, $confusion\_matrix$, $model\_hyperprms$, $model\_perf$, $model\_ref$)\;
     }
     prov.out($'training'$)\;
     prov.close()\;
 
  \end{small}
 \caption{Provenance capture in a training script.}
 \label{alg:alg1}
\end{algorithm}

% \begin{comment}
% \begin{algorithm} 
%  \small{
%  \caption{Provenance capture in a training script.}
%  \label{alg:alg1} 
%  \begin{algorithmic}[1]
%  \renewcommand{\algorithmicrequire}{\textbf{Input:}}
%  \renewcommand{\algorithmicensure}{\textbf{Output:}}
%  \REQUIRE training hyperparameters, input data sets
%  %\ENSURE  out
% %   \STATE first statement
 
%  \textit{import PLLib as prov}
  
%   \STATE prov.init($prospective\_provenance$)
  
%   \textit{...}
  
%   \STATE prov.in($'training'$, $training\_hyperprms$,
  
%   $input\_data\_references$)
%   \FOR {$e = 1$ .. $max\_epochs$}
%   \STATE prov.in($'epoch'$, $e$)
  
%   \textit{...}
  
%   \FOR {$batch\_id$ in $data\_batches$}
%     \STATE prov.in($'batch'$, $batch\_id$)
  
%   \textit{...}
  
%     \STATE prov.out($'batch'$, $loss\_value$)
%   \ENDFOR
  
%   \STATE prov.out($'epoch'$, $confusion\_matrix$,
  
%   $model\_hyperprms$, $model\_perf$, $model\_ref$) 
%   \ENDFOR
%   \STATE prov.out($'training'$)
%   \STATE prov.close()
%  \end{algorithmic} 
%  }
% \end{algorithm}
% \end{comment}

ProvTracker uses prospective provenance data to provide for the tracking by creating the relationships of retrospective provenance data being continuously sent by PLLib, from multiple distributed workflows. ProvTracker gives unique identifiers to every data value captured by the PLLib, so when a data transformation consumes data produced by another, ProvTracker will track such relationship and populate the data graph.  When the data values are data references (\eg{} references to files or identifiers in a database table or any analogous data reference), it creates an edge between the data value and the data store \cite{provlake_escience_2019}. Data transformations that are specific and standard in ML workflows, \eg{} training, validation, and testing are defined beforehand following PROV-ML (\ref{sec_datarepresentation}). ProvTracker also allows users to specify, in the prospective provenance specification, that certain parameters or output values have ML-specific semantics, following PROV-ML, to be stored in the provenance database. Moreover, ProvTracker has work queues to group provenance requests before sending retrospective provenance data to ProvManager. ProvManager is a RESTful service that receives provenance data using PROV-ML vocabulary, and transforms the data into RDF triples (the data model of the DBMS in use by ProvLake in this current implementation) and inserts them in a bulk.

Provenance queries are provided by the PolyProvQueryEngine. 
The characterization (Section \ref{subsec_provanalysis}) and typical queries (\eg{} \queries{}) are used to influence the implementation of parameterized RESTful endpoints using PROV-ML terms. Variations of this endpoint, using terms available in PROV-ML, are used to specify the inputs for the queries. If an endpoint is not implemented for a specific query, users can still write raw queries and submit them to PolyProvQueryEngine directly, which redirects the query to the Prov DBMS. 

\smallskip
\noindent \textbf{Execution Strategies on HPC Clusters.} 
ProvLake uses a microservice architecture to achieve high flexibility when specifying how the components are deployed to reduce performance penalties. 
Fig. \ref{fig:architecture} shows a deployment of ProvLake onto two clusters, one for I/O-intensive workflows like the data processing ones (\textit{Data Proc. Workflows} in the figure---used for the data curation and learning data preparation phases of the lifecycle) and the other for compute-intensive workflows, like the training workflows (for the learning phase). PLLib is the only component in direct contact with the users' workflows running in the clusters, shielding the workflows from possible slowness from other components. To reduce communication cost between the users' workflow and the PLLib, ProvTracker is deployed inside the cluster. To avoid competition (which increases overhead) with the users' workflows, ProvTracker is started on a separate node in the cluster. The other architectural components are deployed externally to the clusters because they are not in direct contact with the PLLib, thus not increasing the communication cost in the workflow scripts. This avoids using extra computing resources only for provenance tracking and analysis, leaving more resources for the users' workflows; and avoids operational work to install more software, such as a DBMS, inside a compute-intensive cluster.

\section{Experimental Validation} \label{sec_exps}

In this section, we provide an experimental validation of ProvLake in support of the \MLCycle{}. 
As execution overhead is a major concern among CSE users, we first present a performance analysis of parallel provenance data capture in Section \ref{subsec_performance_analysis}, 
then we show a running example of which data are captured during the lifecycle of our case study to answer the exemplary queries \queries{} in Section \ref{subsec_usecase}.

\smallskip
\noindent \textbf{Hardware setup.} We use two clusters: a learning cluster, which has 393 Intel and Power8 nodes, each with 24 to 48 CPU cores, 256 to 512 GB RAM, interconnected via InfiniBand, sharing about 3.45 PB in a GPFS, and using in total 946 GPUs (NVIDIA Tesla K40 and K80, each with 2880 and 4992 CUDA cores respectively); and a data processing cluster, which has 12 nodes, each with 128 GB RAM, two Intel CPUs with 40 cores, sharing a GPFS with 24 TB, interconnected via an InfiniBand.

\smallskip
\noindent \textbf{Software setup.} ProvManager, PolyProvQueryEngine, and Prov DBMS are deployed on a virtual Kubernetes cluster with two nodes with 4 vCores, 16 GB RAM each, virtualized on top of the data processing cluster. As in Fig.~\ref{fig:architecture}, the ProvTracker service is started on a separate node on each of the two clusters. ProvLake's services are implemented using Python and deployed with uWSGI  with C++ Cython plugin with multi-process and multi-thread parallelism enabled. ProvManager's queue is set to 50 and ProvTracker threads are set to 120. The workflow scripts of our use case are implemented in Python using multiple libraries, such as to manipulate raw seismic files and for learning (PyTorch V1.1).

\vspace{-1mm}
\subsection{Performance Analysis} \label{subsec_performance_analysis}

\todo[inline]{Justificar as escolhas das metricas de avaliacao.}
In our use case for training an autonomous identifier of geological structures (\emph{c.f.} Sec \ref{subsec_provanalysis}), the learning phase generates a large amount of provenance data at a high frequency to stress ProvLake services. In the deep learning model training, there are two provenance capture calls (for the beginning and end) at each batch iteration, in each learning epoch (\emph{c.f.} Algorithm \ref{alg:alg1}). In this test, each training workflow executes about 35 iterations for each learning epoch and up to 300 epochs, generating about 15,000 provenance capture events per workflow run. ProvTracker runs on one node in the learning cluster with 24 CPU cores, whereas the training workflows run in parallel and distributed on up to 8 nodes, each with 28 Intel CPU cores and 6 GPUs (K80). While running the workflows, PLLib captures data at runtime and sends them to ProvTracker which in turn sends them to ProvManager service deployed externally on the virtual Kubernetes cluster, which finally stores them in the Prov DBMS. A provenance capture overhead analysis of ProvLake using synthetic workloads to highly stress the system and comparison with a competing system has been presented in a previous work \cite{provlake_escience_2019}. Here, we first present a performance analysis testing different settings for provenance analysis, and then a scalability analysis, both using real ML workloads. We measure the overall execution time of the training workflow script, repeating each test at least 10 times and we plot the boxplots of the repetitions and the numeric values used in-text refer to the median of the repetitions.

\begin{figure}
    \centering
    \includegraphics[width=0.85\columnwidth]{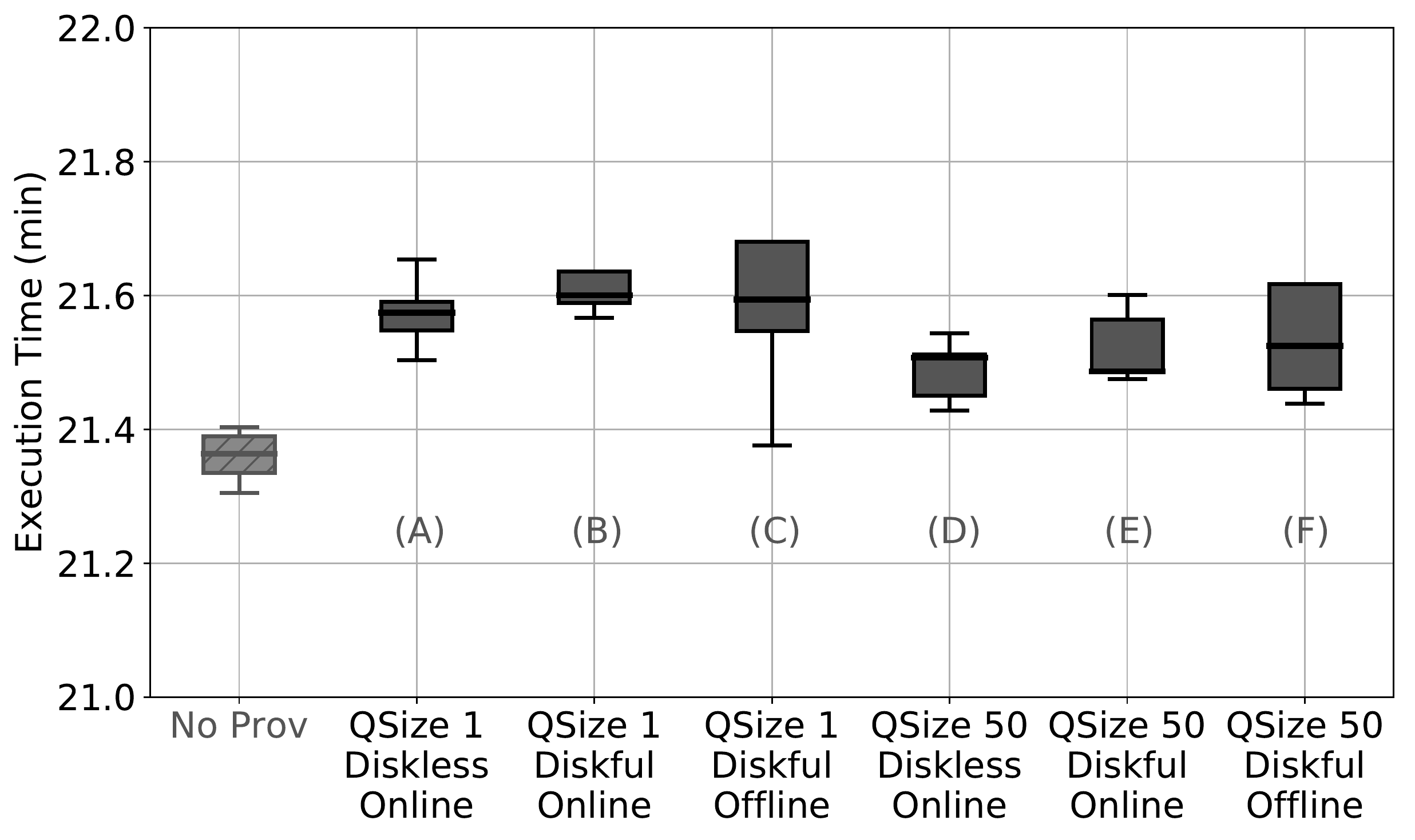}
  \caption{Varying prov. capture. Setting D adds 0.67\% overhead.}
   \label{fig:exp_setting_variation}
   \vspace{-5mm}
\end{figure}

\smallskip
\noindent \textbf{Experiment 1: varying provenance capture settings.} For a baseline, we first execute the training without any provenance capture, then we vary the queue size in PLLib (\ie{} amount of provenance capture requests accumulated in PLLib), diskless vs. diskful (\ie{} saving or not provenance data in a log file on disk), and online vs. offline (\ie{} storing or not provenance data in the DBMS, available for online provenance queries during the execution). As for the training datasets, we use a curated and labeled real seismic dataset using a specific range of seismic slices (corresponding to a regional section of a seismic cube) defined by the model trainer. The results are in Fig.~\ref{fig:exp_setting_variation}, where the fastest result is for Queue Size = 50, Diskless, Online (Setting D). Comparing with the setting with no provenance capture, the added execution overhead in this case is only 8.6 seconds on top of 21.3 minutes, \ie{} 0.67\%, which is considered negligible. 

To analyze the queue size, we compare Settings A--C with D--F and we see larger queues provide faster provenance capture since there is less but larger communication with ProvTracker service. For instance, Setting A is about 7\% slower than D. However, very large queues have drawbacks as they introduce higher latency between the event being captured in the workflow execution and the provenance record being stored in the database, caused by the retention of provenance capture events in PLLib's queue. Nevertheless, for the settings with queue size 50 (D--F), a latency of less than 5 seconds between the actual occurrence of the event and its provenance being registered in the database, available for queries, can be considered near real-time and good enough even for training monitoring. To analyze diskless vs. diskful settings, we compare Setting A with B and C; and D with E and F. Diskless is faster than diskful, as the latter introduces more I/O operations at runtime. However, comparing only the medians, the difference is negligible (less than 0.1\%). Thus, because of a higher fault-tolerance provided by a diskful setting, it may be interesting to append provenance data onto a file on disk, locally in the cluster where the workflow runs. 
%We note, though, that diskful does not imply offline, because we can still send provenance data to the DBMS externally to the cluster while we store provenance data in a file locally, as in settings (B) and (E). 
Similarly, comparing the medians, we observe that the difference between online vs. offline (\eg{} setting B vs. C or E vs. F) is also small, about 1\%. Therefore, despite (D) being the fastest setting, (E) may be preferred because its performance is nearly the same as (D) and it has the advantage of backup storage for provenance data, which is quite important as provenance is used for quality assessment and reproducibility. 
% Removed for space: (especially ProvTracker and ProvManager) 

\begin{figure}
    \centering
    \includegraphics[width=0.85\columnwidth]{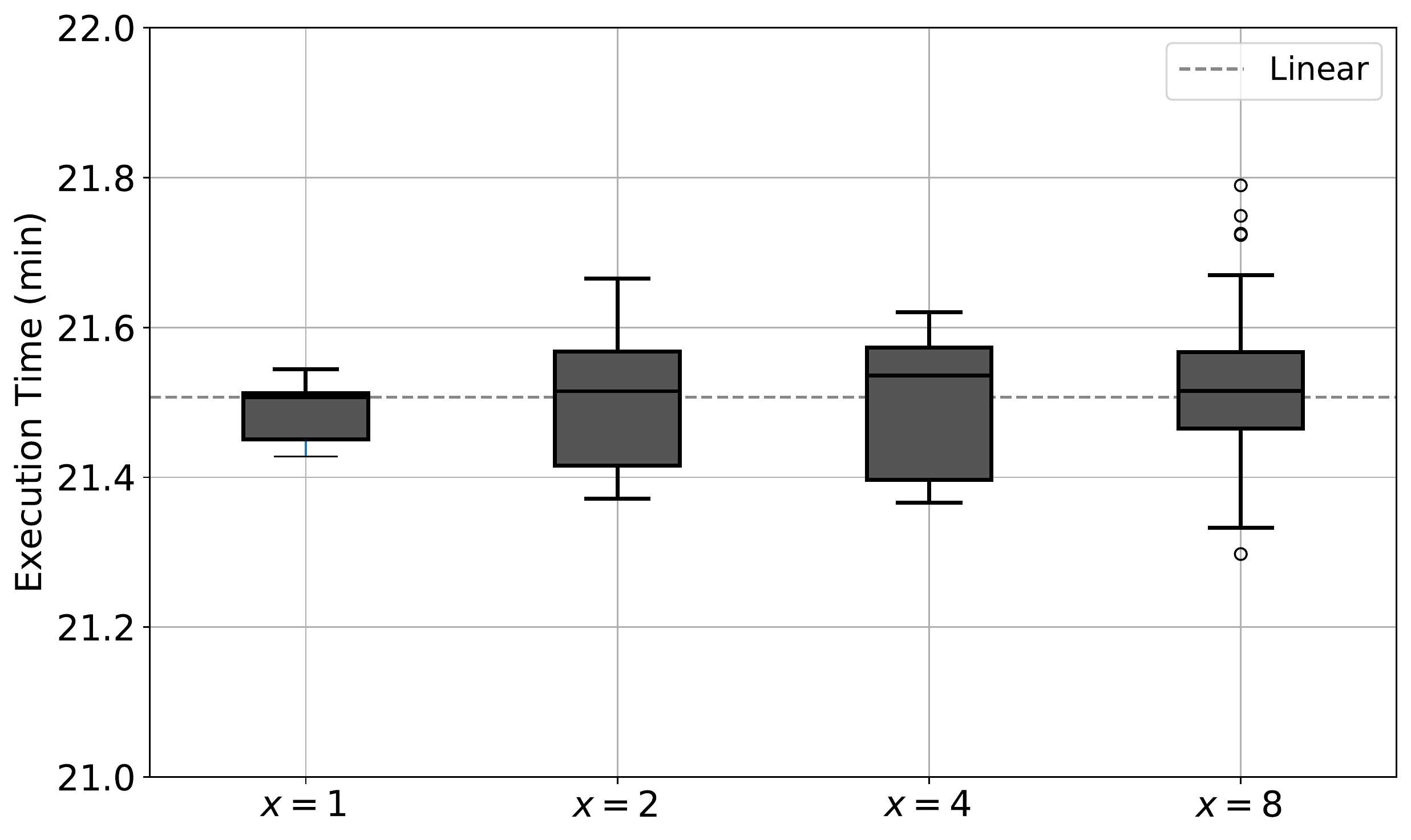}
  \caption{Weak scalability analysis.}
   \label{fig:exp_scalability}
   \vspace{-5mm}
\end{figure}

\smallskip
\noindent \textbf{Experiment 2: scalability analysis.} In this experiment, we want to confirm if the execution strategies on an HPC cluster are keeping the overhead low in a real ML workload, running  multiple training workflows in parallel. We run a weak scalability test by increasing the number of processing units while increasing the data size. We use the fastest setting of the previous experiment (\ie{} D) and the same seismic cube. To set up the training datasets, the trainer selects up to 8 different sets of seismic slices, where each set has the same length (\ie{} nearly the same data size). Thus, for $x \in \{1, 2, 4, 8\}$, there are $x$ workflows running on $x$ nodes in parallel, summing $28x$ Intel CPU cores, $6x$ GPUs, $4992*6x$ CUDA GPU cores, using in total an input dataset with size $x*datasize$, where $datasize$ is the size of a dataset formed by 1 set of seismic slices. The results are in Fig.~\ref{fig:exp_scalability}, where we illustrate the linear scalability as a horizontal line passing through the median of the smallest setting ($x=1$). Ideally, the medians should be near this line. If they are not, it means that ProvTracker is taking too long to answer, caused by high stress in the system due to too many provenance capture requests, adding latency to the training. However, we see that even in the largest setting (\ie{}  $x=8$), the execution time remains close to the linear curve. The boxes remain within a small margin of 0.2 min (or 0.9\% of the $x=1$ median) between 21.4 and 21.6 min, meaning that the system delivers a constant and predictable behavior even at larger scales. We note though that the variance grows with the scale, caused by the larger number of parallel tasks. Therefore, we conclude that at least for this scale (up to 48 K80 GPUs), the provenance capture system delivers good scalability.

\begin{figure*}
  \centering
  \includegraphics[height=5.20cm,keepaspectratio]{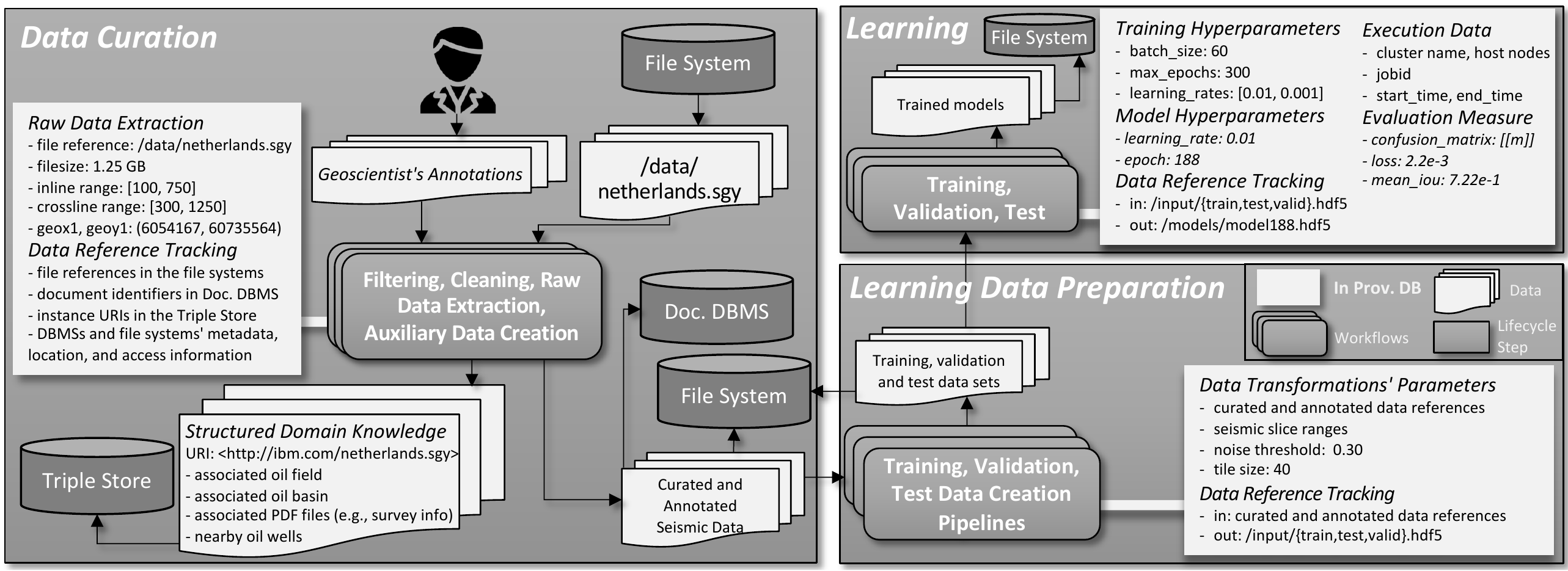}
  \caption{Provenance data tracking in an O\&G use case for the \MLCycle{}.}
   \label{fig:usecase}
   \vspace{-5mm}
\end{figure*}

\vspace{-1.5mm}
\subsection{Use Case Validation} \label{subsec_usecase}

We explain how ProvLake supports queries \queries{} in the O\&G use case, illustrated in Fig.~\ref{fig:usecase} and described in Section~\ref{subsec_provanalysis}. 
As shown in the figure, the phases of the \MLCycle{} are interconnected, as data generated in a phase are consumed in another. Essentially, ProvLake tracks and maintains such interconnections in a provenance data graph as millions of RDF triples (about 30M in total after all models have been trained in this use case) as the chained data transformations in the multiple, distributed workflows composing the inner phases of the major phases of the lifecycle run. The data in the figure are represented as RDF resources, \ie{} instances that extend \codefont{prov:Entity} and PROV-ML specializations. Each of these instances receives a URI, which works as a global identifier throughout the lifecycle. Each trained model generated in the learning phase is represented as an RDF resource, as well as the model hyperparameters of each trained model, the evaluation metrics, and a reference (file path) to the actual model file stored in the file system. Execution data, such as file system metadata, cluster's hostname and node names used in the HPC jobs, job ids in the cluster scheduler, and start and end timestamps of each block of provenance capture events are associated to the trained models in the provenance data graph. 

%removed for space:
% in the form \codefont{(data transformation execution instance, prov:used, input data instance)} and \codefont{(data transformation execution, prov:generated, output data instance)}

Similarly, in the learning data preparation phase, there are several data transformations in a data pipeline that transform the curated and annotated scientific data into training, validation, and test datasets. Each data transformation is parameterized. Parameters specify, for instance, noise filter thresholds, size (in pixels) of tiles which will serve a seismic image classifier, ranges of regions (called seismic slices, \eg{} inline and crossline slices) of the seismic cube that will form the training dataset. Each value of these parameters, the name of the transformation, execution data, and data references to input and output data physically stored in the data stores are captured and represented in ProvLake's provenance data graph. For the data curation phase, ProvLake captures provenance when the data-intensive scripts that clean, filter, and create auxiliary data run. When processing raw scientific files, important data which will help to answer the queries are extracted, such as geographic coordinates embedded as metadata in raw SEG-Y seismic files (represented as a \codefont{netherlands.sgy} in the figure), associated to the file's URI, and stored in the provenance database. Yet, geoscientists input important annotations into some of those scripts including associated oil fields, basins, oil wells, and pieces of texts from PDF documents with survey information related to the geological data acquisition process. These annotations are stored in a domain-specific database, externally to the provenance database, stored in a Triple Store. In this case, ProvLake's ability to keep track of data distributed in multiple stores helps to maintain the data relationships between the raw files in the file system and the structured knowledge stored in another database. Auxiliary data, such as polygons of the seismic cube are stored in the Document DBMS, and the data references are similarly tracked and related to the raw files. Other data, such as implementation details, software name and version, are captured and stored in the provenance database, following the PROV-ML, but, for simplicity, we do not show them in the figure. As the data and their relationships are properly tracked while the workflows execute, ProvLake enables answering online, offline, intra- and inter-training provenance queries to analyze ML data, domain-specific data, and execution data throughout the phases of the lifecycle, exemplified by the queries \queries{}.

To submit queries, the user sends a GET or POST request to one of PolyProvQueryEngine's endpoints. Then, PolyProvQueryEngine sends requests to ProvManager. Most of the queries are answered with simple graph traversals using standard SPARQL features. For instance, to answer Q1, the user provides a trained model URI (generated in the learning phase) and the query should traverse in the provenance data graph backward until the raw seismic file's URI (processed in the data curation phase). To return the geographic coordinates and number of seismic slices, the query uses the extracted data related to the seismic file. To return the oil basin and oil field information, the query retrieves data from the resource, in the Triple Store, that represents structured knowledge about the seismic file. For Q2 and Q6, similar graph traversal is executed. Other queries require analytical operators, such as Q3, which requires finding the trained model with least (using \codefont{min()} native SPARQL operator) loss, and returning its hyperparameters. Q4 and Q5 make use of execution data to provide basic statistics (\codefont{min(), max(), avg()} operators) about the execution time of training iterations.

\vspace{-1mm}
\section{Related Work} \label{sec_related_work}

Some works have addressed provenance tracking in the ML context \cite{miao_towards_2017,zhang_diagnosing_2017,kumar_model_2016,zaharia_accelerating_2018}. However, they are mainly focused on the learning and learning data preparation phases, failing to trace back from the trained models until the raw domain-specific data curated in workflows in the curation phase. 
These solutions often come with one single system to manage execution, data, and provenance of the whole lifecycle, but in order to do so, users need to develop their workflows in such a system. 
Although it is a good fit for simple projects (\eg{} the same user designs ML models, curates, prepares the data and trains the ML models), it is not for CSE, which is considerably more complex and heterogeneous. 
It is unrealistic to expect that all phases, their execution, and the processed data will be managed by one single system.
Alternatively, provenance tracking systems \provcapturesystems{} can be coupled to a CSE workflow, providing provenance support while not significantly changing the way CSE users develop their applications. However, these solutions fail to track the interconnections between workflows and fail to track data processed in multiple heterogeneous stores. Also, some of them \cite{komadu_suriarachchi_big_2018} add high provenance capture overhead, preventing their adoption in CSE. Finally, none of them has a provenance data representation capable of representing ML-specific and domain-specific data, as we propose with PROV-ML, with extensions of W3C PROV \cite{W3CPROV} and MLS \cite{W3CML}.

On new provenance data representations for ML, some works addressed the gap between the experiments of a ML workflow execution and a standard representation to provide reproducible experiments~\cite{esteves2015, publio2018, W3CML}. Esteves~\etal{}~\cite{esteves2015} introduce W3C PROV-compliant provenance in these workflows in ML. They provide a machine-readable vocabulary and a common schema for reproducibility in various frameworks and workflow systems. However, it lacks details of the ML phases itself. Publio~\etal{}~\cite{publio2018} present a new ML data representation based on MEX vocabulary \cite{esteves2015} to improve processes on ML workflows. Nonetheless, they lack a explicit separation between prospective and retrospective provenance, limiting provenance data understanding. Moreover, these works are focused on the learning phases of the lifecycle, whereas the interconnections with workflows in prior phases, like for data curation, are not provided. Finally, none of these solutions has a provenance tracking system as we are proposing.

\section{Conclusions} \label{sec_conclusion}

In this work, we addressed the problem of tracking the data transformations in the ML lifecycle, focusing on CSE. We showed that heterogeneity in several dimensions, including different human expertise, workflows, data stores, execution machines, among others, adds a significant complexity that must be addressed to support provenance tracking in the ML lifecycle; end-to-end from raw scientific data files to trained models. To the best of our knowledge, this is the first work that characterizes provenance as an essential aspect to be managed for the track of data in the ML lifecycle in CSE.

Although existing provenance tracking solutions that can be coupled with workflows contribute with the flexibility needed in CSE projects, they fail to support the heterogeneous nature of the lifecycle. After the practical experience of extending ProvLake for the lifecycle, we draw the following lessons: 
\begin{enumerate}[label=(\roman*),wide,labelindent=3mm] 
%,wide, labelwidth=!, labelindent=0pt]
    \item The characterization of provenance in the lifecycle allows for an understanding of the different needs of different persona as it drives the provenance tracking to answer key online and offline, intra- and inter-training provenance queries capable of analyzing, in an integrated way, ML data, domain-specific data, and execution data, throughout the data curation, data preparation and learning phases of the lifecycle. We observed that the data curation step, which is often neglected by ML systems, is the most complex part in CSE and needs to be addressed carefully for provenance analysis.

    \item In CSE, it is necessary to integrate provenance from multiple workflows that process domain-specific data in the data curation phase and ML data in the learning phases of the ML lifecycle; otherwise, important data are not tracked properly. In practice, this is often done manually, which is time consuming and error prone. To achieve this integration, it was key to create a representation that leverages ML and domain-specific data. Therefore, we created PROV-ML, which is compliant to W3C definitions, namely PROV and ML Schema. We hope such representation can be adopted by other systems in this area.
    
    \item Architectural design decisions, such as a microservice architecture and a lightweight provenance capture library (with less than 1\% of overhead), are essential for efficient tracking with comprehensive provenance analysis. We observed this through a real O\&G case running on a testbed of 48 GPUs.
\end{enumerate}

\vspace{-2.5mm}
\section*{Acknowledgment}
\begin{footnotesize}
\noindent We thank Marcelo Costalonga and Daniela Szwarcman from IBM Research Brazil for their help, and the WORKS19 anonymous reviewers for their valuable feedback. This work was partially funded by CNPq, FAPERJ, and Inria Associated Team SciDISC.

\end{footnotesize}

\vspace{-1.0mm}
\bibliography{references}
\bibliographystyle{IEEEtran.bst}

\end{document}